\documentclass[           % Format : preprint, twocolumn
               showpacs,            % Pacs : showpacs, noshowpacs
               preprintnumbers,     % Preprint: preprintnumbers,
               aps,                 % Society: ...
               prd,                 % Journal Style : pra, prb, prc, prd, pre,
               superscriptaddress,      % Affiliation (Title) : groupedaddress,
               nofootinbib,         % Footnote: footinbib, nofootinbib
               tightenlines,        % Remove additional spaces in a line
               floats,floatfix      % Floating pictures and tables
               ]{revtex4}

\usepackage{amsmath}
\usepackage{amssymb}
\usepackage{graphicx}
\newcommand{\Real}{\mathbb{R}}

\begin{document}

% ----->   TITLE   <-----

\title{Instability of wormholes supported by a ghost scalar field. 
I. Linear stability analysis}

% ----->   AUTHORS   <-----

\author{J. A. Gonz\'alez}
\affiliation{Instituto de F\'{\i}sica y Matem\'{a}ticas, Universidad
              Michoacana de San Nicol\'as de Hidalgo. Edificio C-3, Cd.
              Universitaria, A. P. 2-82, 58040 Morelia, Michoac\'{a}n,
              M\'{e}xico.}

\author{F. S. Guzm\'an}
\affiliation{Instituto de F\'{\i}sica y Matem\'{a}ticas, Universidad
              Michoacana de San Nicol\'as de Hidalgo. Edificio C-3, Cd.
              Universitaria, A. P. 2-82, 58040 Morelia, Michoac\'{a}n,
              M\'{e}xico.}

\author{O. Sarbach}
\affiliation{Instituto de F\'{\i}sica y Matem\'{a}ticas, Universidad
              Michoacana de San Nicol\'as de Hidalgo. Edificio C-3, Cd.
              Universitaria, A. P. 2-82, 58040 Morelia, Michoac\'{a}n,
              M\'{e}xico.}

% --->   DATE

\date{\today}

% ----->   ABSTRACT   <-----

\begin{abstract}
We examine the linear stability of static, spherically symmetric
wormhole solutions of Einstein's field equations coupled to a massless
ghost scalar field. These solutions are parametrized by the areal
radius of their throat and the product of the masses at their
asymptotically flat ends. We prove that all these solutions are
unstable with respect to linear fluctuations and possess precisely one
unstable, exponentially in time growing mode. The associated time
scale is shown to be of the order of the wormhole throat divided by
the speed of light. The nonlinear evolution is analyzed in a
subsequent article.
\end{abstract}

% ----->   PACS

\pacs{04.20.-q, 04.25.-g, 04.40.-b} 
% 04.20.-q: Classical GR
% 04.20.Ex: Initial value problem, existence and uniqueness of solutions
% 04.25.-g: Approximation methods; equations of motion
% 04.25.D-: Numerical relativity
% 04.40.-b: Self-gravitating systems, continuous media and 
%           classical fields in curved spacetime
% 04.70.-s: Physics of black holes
% 97.60.Lf: Astronomy: Late stage of star evolution: black holes

% ----->   MAKETITLE   <-----

\maketitle

% ----------------------
% ----->   BODY   <-----
% ----------------------

% ----->     INTRODUCTION

\section{Introduction}

Einstein's general theory of relativity leads to spectacular
predictions like the big bang in the early universe and the existence
of black holes and gravitational radiation, all of which are subject
to mainstream research in astrophysics and cosmology. More speculative
predictions of Einstein's theory are traversable wormhole geometries
\cite{mMkT88} consisting of two asymptotically flat ends connected by
a common throat through which time-like test particles might
travel. Similar constructions can be obtained by removing two holes
from a given three-manifold and connecting one with another by a
handle, in which case wormholes have been proposed for interstellar
travel or building time machines \cite{mMkTyU88,vFiN90}.

However, as a consequence of the topological censorship theorem
\cite{jFkSdW93} wormholes must be supported by matter fields which are
``exotic'' in the sense that they violate the averaged null energy
condition. In particular, this implies the violation of the weak
energy condition which means that some time-like observers measure a
negative energy density. On the other hand, it has been shown recently
\cite{mVsKnD03} that it is possible to construct wormholes where the
violation of the averaged null condition can be made arbitrarily
small. Therefore, such wormholes could in principle be supported by
quantum fields since quantum effects can lead to violations of the
energy conditions.

In this work we assume the existence of stationary wormhole geometries
and analyze their stability with respect to small initial
perturbations. If stable, the time evolution of such a perturbation
may deform the wormhole throat by a small amount, but for large times
the perturbation decays to zero and the equilibrium configuration is
recovered. If on the other hand the wormhole is unstable, the time
evolution may induce large deformations of the throat and in this case
it is possible that the wormhole is destroyed making it unsuitable for
interstellar travel or building time machines unless the time scale
associated to the instability is large enough.

Clearly, the notion of stability depends on the precise matter model
that is being considered. In this work, for simplicity, we focus on a
massless ghost scalar field. By a ghost field we mean one which
violates the dominant energy condition. A massless ghost scalar field
consists of a minimally coupled, massless scalar field whose kinetic
energy has a reversed sign. Such a field violates the null, and as a
consequence also the weak and dominant energy conditions. We analyze
the stability of static, spherically symmetric wormhole geometries
supported by such fields. We prove in this paper that all such
solutions are unstable with respect to linear perturbations and
possess precisely one unstable mode which grows exponentially in
time. Furthermore, we show that the associated time scale is of the
order of the areal radius of the wormhole throat divided by the speed
of light. In particular, this means that a wormhole whose throat has
an areal radius smaller than $1km$, say, decays in a few $\mu s$.

Wormholes supported by scalar fields have been considered before (see,
for instance,
\cite{kBsG02,kBsG04,kBaS07,cA02,hSsH02,sSsK04,tMdN06,eDtZ07}). In
\cite{kBsG02,kBsG04,kBaS07} wormholes with a nonminimally coupled
scalar field are shown to be unstable with respect to linear
fluctuations. In \cite{cA02} wormholes with a massless, ghost scalar
field are constructed. These wormholes are static, spherically
symmetric and connect two asymptotic ends which either have both
vanishing ADM masses or have ADM masses of opposite signs. Based on a
linear stability analysis for the zero mass case it was claimed in
\cite{cA02} that the wormholes with sufficiently small ADM masses are
stable. However, a numerical simulation performed in \cite{hSsH02}
indicates that small perturbations of the zero mass wormholes do not
decay under nonlinear evolution and may cause the wormhole to explode
or to collapse. Here, we point out that the linear stability analysis
of \cite{cA02} is incomplete in the sense that it only shows stability
of the zero mass wormholes for a restricted class of perturbations
where the areal radius is unperturbed and the {\em perturbed scalar
field vanishes on the throat}. However, such perturbations are
artificial in some sense because one could imagine perturbing the
scalar field by a small ingoing pulse which is supported away from the
throat at some time $t=0$, say. As $t$ grows, this pulse travels
towards the throat and since the wormhole metric is everywhere regular
one expects it to reach and cross the throat at some finite time. On
the other hand, requiring that the perturbed scalar field vanishes at
the throat corresponds to placing a mirror at the throat which
reflects the scalar pulse. As we show in this article, the zero mass
wormholes are {\em linearly unstable} in the absence of such a mirror,
i.e. if one allows perturbations which are not required to vanish at
the throat. Furthermore, we generalize the stability analysis to the
static, spherically symmetric wormholes with nonvanishing ADM masses
and prove that they are linearly unstable as well. In a subsequent
paper we analyze the nonlinear evolution of small initial
perturbations by numerical means and show that depending on the
details of the initial perturbation the wormhole either expands or
collapses to a Schwarzschild black hole.

This work is organized as follows. In
section~\ref{Sec:StaticSolutions} we review the static, spherically
symmetric wormhole configurations which are supported by a ghost
scalar field. In section~\ref{Sec:LinStabAnal} we perform the linear
stability analysis. We do so by first generalizing the master equation
obtained in \cite{cA02} to massive wormholes. This master equation
turns out to be singular at the throat for both massive and massless
wormholes. As a consequence, it requires the solution to decay
sufficiently fast as the throat is approached. We then transform the
singular master equation in a regular one and show that the new master
equation possesses an unstable mode which grows exponentially in
time. The associated time scale is computed in
section~\ref{Sec:TimeScale} by a numerical shooting method. Finally,
conclusions are drawn in section~\ref{Sec:Conclusions}.

%%%%%%%%%%%%%%%%%%%%%%%%%%%%%%%%%%%%%%%%%%%%%%%%%%%%%%%%%%%%%%%%%%%%%%%
\section{Static, spherically symmetric wormholes supported by a 
ghost scalar field}
\label{Sec:StaticSolutions}
%%%%%%%%%%%%%%%%%%%%%%%%%%%%%%%%%%%%%%%%%%%%%%%%%%%%%%%%%%%%%%%%%%%%%%%

We consider a gravitational field $g$ which is coupled to a massless
ghost scalar field $\Phi$. The action is
\begin{displaymath}
S[g,\Phi] = \frac{1}{16\pi G}
\int\left( -R + \kappa\nabla^\mu\Phi\cdot\nabla_\mu\Phi \right)
\sqrt{-g}\, d^4 x,
\end{displaymath}
where $R$ and $\nabla$ denote the Ricci scalar and the covariant
derivate, respectively, associated to $g$, and $\kappa = -8\pi G$ with
$G$ Newton's constant. Notice that $\kappa$ is negative which is the
reason for calling the scalar field ``ghost''. The corresponding field
equations are
\begin{eqnarray}
R_{\mu\nu} &=& \kappa\; \nabla_\mu\Phi \cdot \nabla_\nu\Phi, 
\label{Eq:Einstein}\\
0 &=& \nabla^\mu\nabla_\mu\Phi,
\label{Eq:KleinGordon}
\end{eqnarray}
where $R_{\mu\nu}$ is the Ricci tensor associated to $g$.

For a spherically symmetric field, local coordinates
$t,x,\vartheta,\varphi$ can be chosen such that
\begin{equation}
g = -e^{2d} dt^2 + e^{2a} dx^2 
 + e^{2c}\left( d\vartheta^2 + \sin^2\vartheta\; d\varphi^2 \right),
\label{Eq:SphericalMetric}
\end{equation}
where the functions $d = d(t,x)$, $a = a(t,x)$ and $c = c(t,x)$, as
well as the scalar field $\Phi = \Phi(t,x)$ depend only on the time
coordinate $t$ and the spatial coordinate $x$. We are interested in
traversable wormhole geometries which consist of a throat connecting
two asymptotically flat ends at $x\to \infty$ and $x\to -\infty$,
respectively. This means that the areal radius $r = e^c$ is strictly
positive and proportional to $|x|$ for large $|x|$ and that the
two-manifold $(\tilde{M},\tilde{g}) = (\Real^2,-e^{2d} dt^2 + e^{2a}
dx^2)$ is regular and asymptotically flat at $x\to \pm\infty$.

For the spherically symmetric ansatz (\ref{Eq:SphericalMetric}) the
Einstein equations (\ref{Eq:Einstein}) yield the Hamiltonian
constraint ${\cal H} := -e^{a-d}( R_{tt} + e^{2d} R/2) +
\kappa(e^{a-d}\Phi_t^2 + e^{d-a}\Phi_x^2)/2 = 0$, the momentum
constraint ${\cal M} := -(R_{xt} - \kappa\Phi_t\Phi_x) = 0$ and the
evolution equations $e^{d-a}( R_{xx} - \kappa\Phi_x^2 ) + {\cal H} =
0$ and $R_{\vartheta\vartheta} = R_{\varphi\varphi} = 0$. This and the
wave equation (\ref{Eq:KleinGordon}) yields the evolution equations
\begin{eqnarray}
\partial_t\left( e^{a-d} a_t \right) - \partial_x\left( e^{d-a} d_x \right)
 - e^{a-d} c_t^2 + e^{d-a} c_x^2 - e^{a+d-2c} 
 &=& -\frac{\kappa}{2}\left[ e^{a-d}\Phi_t^2 - e^{d-a}\Phi_x^2 \right],
\label{Eq:Ev1}\\
\partial_t\left( e^{a-d+2c} c_t \right) 
 - \partial_x\left( e^{d-a+2c} c_x \right) &=& -e^{a+d},
\label{Eq:Ev2}\\
\partial_t\left( e^{a-d+2c} \Phi_t \right) 
 - \partial_x\left( e^{d-a+2c} \Phi_x \right) &=& 0,
\label{Eq:Ev3}
\end{eqnarray}
and the constraints are
\begin{eqnarray}
{\cal H} &:=& e^{d-a}\left[ 2c_{xx} + (3c_x - 2a_x)c_x \right] 
 - e^{a-d} c_t(2a_t + c_t) - e^{a+d-2c}
 + \frac{\kappa}{2}\left[ e^{a-d}\Phi_t^2 + e^{d-a} \Phi_x^2 \right] = 0,\\
{\cal M} &:=& 2c_{tx} + 2c_t c_x - 2d_x c_t - 2a_t c_x 
 + \kappa\, \Phi_t \Phi_x = 0.
\end{eqnarray}
Here, the subscript $t$ and $x$ refer to the derivatives with respect
to $t$ and $x$, respectively.

For a static configuration, the scalar field $\Phi$ and the metric
coefficients $d$, $a$ and $c$ are independent of $t$. In this case the
field equations can be integrated analytically \cite{hE73,kB73}. In
order to see this we adopt a gauge where $d=-a$ in which case the
field equations reduce to
\begin{equation}
\left[ e^{2(c-a)} \right]_{xx} = 2, \qquad
\left[ e^{2(c-a)} c_x \right]_x = 1, \qquad
\left[ e^{2(c-a)}\Phi_x \right]_x = 0,
\label{Eq:Static1}
\end{equation}
and
\begin{equation}
c_x^2 - 2 a_x c_x - e^{2(a-c)} - \frac{\kappa}{2}\, \Phi_x^2 = 0.
\label{Eq:Static2}
\end{equation}
The first equation has the general solution $e^{2(c-a)} = x^2 +
2\alpha_1 x + \alpha_0$ with two constants $\alpha_0$ and
$\alpha_1$. By a suitable translation of the coordinate $x$ it is
always possible to obtain $\alpha_1=0$. Furthermore, since we are
interested in wormhole geometries with the properties described below
equation~(\ref{Eq:SphericalMetric}) we need $e^{2(c-a)} > 0$ for all
$x\in\Real$. Therefore, we have $e^{2(c-a)} = x^2 + b^2$ with some
strictly positive constant $b > 0$. The second and third equations in
(\ref{Eq:Static1}) have the solutions
\begin{equation}
c = \frac{1}{2}\log(x^2 + b^2) 
 - \gamma_1\arctan\left( \frac{x}{b} \right) - \gamma_0\; ,\qquad
\Phi = \Phi_1\arctan\left( \frac{x}{b} \right) + \Phi_0\; ,
\label{Eq:StaticWormHole1}
\end{equation}
with constants $\gamma_0$, $\gamma_1$, $\Phi_0$, $\Phi_1$. It follows
that
\begin{equation}
d = -a = \gamma_1\arctan(x/b) + \gamma_0\; ,
\label{Eq:StaticWormHole2}
\end{equation}
and the metric reads
\begin{equation}
g = -e^{2\gamma_1\arctan(x/b) + 2\gamma_0} dt^2
 + e^{-2\gamma_1\arctan(x/b) - 2\gamma_0}\left[
 dx^2 + (x^2 + b^2)\left( d\vartheta^2 + \sin^2\vartheta\; d\varphi^2 \right)
 \right].
\label{Eq:WormholeMetric}
\end{equation}
Notice that the constant rescaling $t \mapsto \exp(-\Omega) t$,
$x\mapsto \exp(\Omega) x$, $b \mapsto \exp(\Omega) b$, $\gamma_0
\mapsto \gamma_0 + \Omega$, $\gamma_1 \to \gamma_1$ with $\Omega$ a
nonvanishing constant leaves the metric unchanged. So for example, we
can rescale the coordinates such that $\gamma_1\pi + 2\gamma_0 = 0$ in
which case $e^{2a}\to 1$ and $e^{2c}/x^2\to 1$ as $x\to
+\infty$. Alternatively, we can rescale the coordinates such that
$-\gamma_1\pi + 2\gamma_0 = 0$ in which case $e^{2a}\to 1$ and
$e^{2c}/x^2\to 1$ as $x\to -\infty$. With these observations in mind
we see that the metric (\ref{Eq:WormholeMetric}) has indeed two
asymptotically flat ends at $x\to +\infty$ and $x\to -\infty$,
respectively. Finally, equation~(\ref{Eq:Static2}) yields the relation
\begin{equation}
-\kappa\Phi_1^2 = 2(1 + \gamma_1^2)
\label{Eq:WormholeParamConstr}
\end{equation}
between the parameters $\gamma_1$ and $\Phi_1$, where we recall that
$\kappa = -8\pi G$ is negative.

Summarizing, we obtain the bi-parametric family of wormhole solutions
described by equations (\ref{Eq:WormholeMetric}) and
(\ref{Eq:StaticWormHole1},\ref{Eq:StaticWormHole2}). This family of
solutions has been obtained in \cite{hE73,kB73}, and in \cite{cA02} in
the context of wormholes. The solutions can be parametrized by the
constants $B := b e^{-\gamma_0} > 0$ and $\gamma_1 \geq 0$ which are
invariant with respect to the rescaling $b \mapsto \exp(\Omega) b$,
$\gamma_0 \mapsto \gamma_0 + \Omega$ discussed above. $\Phi_1$ is
determined (up to its sign) by the constraint
(\ref{Eq:WormholeParamConstr}). (The sign of $\Phi$ is not important
since the field equations are invariant with respect to the
transformation $\Phi \mapsto -\Phi$. Similarly, the parameter $\Phi_0$
has no physical meaning since only the gradient of $\Phi$ appears in
the equations. Furthermore, it is sufficient to consider nonnegative
values for $\gamma_1$ since the solution is invariant with respect to
a change of sign of both $\gamma_1$ and $x$.) In order to relate the
parameters $B$ and $\gamma_1$ to physical quantities we first remark
that the areal radius,
\begin{displaymath}
r = e^c = B\sqrt{1 + \left(\frac{x}{b}\right)^2}\; 
e^{-\gamma_1\arctan(x/b)},
\end{displaymath}
has a global minimum at $x = x_{throat} = \gamma_1 b$ since $c_x = (x
- \gamma_1 b)/(x^2 + b^2)$ and $\lim\limits_{x\to\pm\infty} (r/|x|) =
\exp(-\gamma_0 \mp \gamma_1\pi/2)$. Next, we compute the Misner-Sharp
mass function \cite{cMdS64}. For the spherically symmetric spacetime
metric given by equation~(\ref{Eq:SphericalMetric}) it is defined by
\begin{displaymath}
m(t,x) := \frac{r}{2}\left[ 1 - \tilde{g}(dr,dr) \right] 
 = \frac{e^c}{2}\left[ 1 + e^{2(c-d)} c_t^2 - e^{2(c-a)} c_x^2 \right].
\end{displaymath}
For the static solutions described by
equations~(\ref{Eq:StaticWormHole1},\ref{Eq:StaticWormHole2}) this
yields
\begin{displaymath}
m(x) = \frac{r}{2}\left[ 1 - \frac{(x - \gamma_1 b)^2}{x^2 + b^2} \right].
\end{displaymath}
The ADM masses of the two asymptotically flat ends can be computed by
considering the asymptotic values $m_{\pm\infty} :=
\lim\limits_{x\to\pm\infty} m(x)$ of the mass function, which yields
$m_\infty = B\gamma_1\exp(-\gamma_1\pi/2)$ and $m_{-\infty} =
-B\gamma_1\exp(\gamma_1\pi/2)$. Notice also that at the throat,
$m(x_{throat}) = r_{throat}/2$.

Therefore, the parameters $B$ and $\gamma_1$ determine the ADM masses
of the two asymptotic ends, $m_\infty$ and $m_{-\infty}$, and the
areal radius of the throat, $r_{throat} = B\sqrt{1 + \gamma_1^2}\;
e^{-\gamma_1\arctan(\gamma_1)}$. Conversely, the areal radius of the
throat and the product of the asymptotic masses uniquely determine the
parameters $B$ and $\gamma_1$. The particular case $\gamma_1 = 0$
yields a zero mass wormhole, $m_{\infty} = m_{-\infty} = 0$, in all
other cases the asymptotic masses are nonzero and have opposite signs.

Before concluding this section, we compute the violation of the
averaged null energy condition according to \cite{mVsKnD03}, which is
given by the volume integrals of $\rho + p_{\hat{r}}$ over the two
asymptotically flat ends,
\begin{displaymath}
I_{\pm} := 4\pi \int\limits_{x_{throat}}^{\pm \infty}
 (\rho + p_{\hat{r}}) r^2 r_x dx,
\end{displaymath}
where $\rho$ and $p_{\hat{r}}$ are, respectively, the energy density
and the radial pressure as measured by static observers. Since $\rho
= p_{\hat{r}}$ for a scalar field and since the Hamiltonian constraint
implies that
\begin{displaymath}
(2m)_x = \frac{\kappa}{2} e^{3c-2a} c_x\Phi_x^2 = |\kappa| \rho\, r^2 r_x\; ,
\end{displaymath}
for static solutions we obtain
\begin{equation}
I_\pm = \frac{8\pi}{|\kappa|} (2m_{\pm\infty} - r_{throat}) 
 = \frac{8\pi}{|\kappa|} r_{throat}\left[ 
 \pm 2\frac{\gamma_1}{\sqrt{1 + \gamma_1^2}} 
 e^{\gamma_1[\arctan(\gamma_1) \mp \pi/2]} - 1 \right].
\end{equation}
We have $I_- \leq I_+ < 8\pi|\kappa|^{-1} r_{throat}(2e^{-1} - 1) <
0$, so both integrals are strictly negative as expected from the
violation of the averaged null energy condition. However, we also see
that both integrals can be made arbitrarily small by fixing $\gamma_1$
and letting $r_{throat} \to 0$.

%%%%%%%%%%%%%%%%%%%%%%%%%%%%%%%%%%%%%%%%%%%%%%%%%%%%%%%%%%%%%%%%%%%%%%%
\section{Linear stability analysis}
\label{Sec:LinStabAnal}
%%%%%%%%%%%%%%%%%%%%%%%%%%%%%%%%%%%%%%%%%%%%%%%%%%%%%%%%%%%%%%%%%%%%%%%

In this section we analyze the linear stability of the two-parameter
family of static wormhole solutions discussed in the previous
section. In order to do so, we consider small perturbations of the
form
\begin{displaymath}
\Phi(\lambda) = \Phi + \lambda\delta\Phi + {\cal O}(\lambda^2),
\end{displaymath}
where $\Phi$ is the background solution, and where
\begin{displaymath}
\delta\Phi := \left. \frac{d}{d\lambda} \Phi(\lambda) \right|_{\lambda=0}
\end{displaymath}
denotes the variation of $\Phi$. The same applies to the other fields
$d$, $a$ and $c$. Since there are no dynamical degrees of freedom for
a spherical symmetric gravitational field we expect to be able to
describe linear fluctuations by a single master equation for the
linearized scalar field $\delta\Phi$. In fact this is known
\cite{oBmHnS96} to be the case for a large class of matter fields when
considering linear fluctuations of static configurations.

However, as we will see in Sec. \ref{Sec:LinStabMasterPhi} the
derivation of the master equation for $\delta\Phi$ requires a gauge
where the areal radius is unperturbed, i.e. where $\delta r =
0$. Since $r = e^c$ is a scalar field on the two-manifold $\tilde{M}$
its linearization $\delta r$ transforms as
\begin{equation}
\delta r \mapsto \delta r + \pounds_\xi r = \delta r + \xi^x r_x
\end{equation}
with respect to infinitesimal coordinate transformations on
$\tilde{M}$ which are parametrized by the vector field $\xi$. If $r_x
\neq 0$ it is always possible to achieve the gauge $\delta r = 0$ by
an appropriate choice of $\xi^x$. However, for wormhole topologies,
$r_x = 0$ at the throat; hence it is not possible to set $\delta r =
0$ everywhere unless we restrict ourselves to perturbations which hold
the areal radius of the throat (and any other surface of critical $r$)
fixed.

An alternative approach which does not require the gauge $\delta r =
0$ is to replace $\delta\Phi$ with the linear combination
\begin{equation}
\Psi := e^c\left( \delta\Phi - \frac{\Phi_x}{c_x} \delta c \right),
\label{Eq:DefPsi}
\end{equation}
which is invariant with respect to infinitesimal coordinate
transformations on $\tilde{M}$. In Sec. \ref{Sec:LinStabMasterPhi} we
show that the linear perturbations can in fact be described by a
master equation which has the form of a wave equation for $\Psi$ with
potential $V$. However, we will see that this potential {\em diverges}
as $x\to x_{throat}$ approaches the throat. This property is not
entirely surprising since for $\delta c \neq 0$ the gauge-invariant
quantity $\Psi$ is not well-defined at the throat where $\Phi_x/c_x$
diverges.

In Sec. \ref{Sec:LinStabMasterNew} we transform the singular master
equation for $\Psi$ into a new master equation which also has the form
of a wave equation with a new potential $W$ which is everywhere
regular, thus facilitating the stability analysis. Furthermore, we
prove in Sec. \ref{Sec:LinStabBoundState} that the new equation
possesses a unique unstable mode which grows exponentially in
time. This mode is then shown in \ref{Sec:LinStabUnstable} to give
rise to an exponentially growing solution of the linearized field
equations which is everywhere regular, thus proving the linear
instability of the static, spherically symmetric wormholes supported
by a massless ghost scalar field.

\subsection{Master equation for $\Psi$}
\label{Sec:LinStabMasterPhi}

Here, we derive the master equation for $\Psi$ which is valid for
points away from the throat. In order to do so, we first consider the
linearized Hamiltonian and momentum constraints which read
\begin{eqnarray}
\delta( e^{a-d} {\cal H}) &=& 2\delta c_{xx} + (6c_x - 2a_x)\delta c_x 
 - 2c_x\delta a_x - 2(\delta a - \delta c)e^{2(a-c)} + \kappa\Phi_x\delta\Phi_x
 = 0,\\
\delta {\cal M} &=& 2\delta c_{tx} + 2(c_x - d_x)\delta c_t - 2c_x\delta a_t 
 + \kappa\, \Phi_x\delta\Phi_t = 0.
\end{eqnarray}
With the help of the background equations $(e^{d-a+2c} d_x)_x = 0$,
$(e^{d-a+2c} c_x)_x = e^{a+d}$ and $(e^{d-a+2c}\Phi_x)_x = 0$, one can
show that these equations possess the first integral
\begin{equation}
\delta c_x + (c_x - d_x)\delta c - c_x\delta a 
 = -\frac{\kappa}{2}\Phi_x\delta\Phi + \sigma\, e^{a-d-2c},
\label{Eq:FirstIntegral}
\end{equation}
where $\sigma$ is a constant whose meaning is explained below.

Next, the linearization of equations~(\ref{Eq:Ev2},\ref{Eq:Ev3})
yields
\begin{eqnarray}
\delta c_{tt} - e^{d-a-2c}\partial_x\left( e^{d-a+2c}\delta c_x \right)
 &=& c_x e^{2(d-a)} \partial_x(\delta d - \delta a + 2\delta c)
  - 2 e^{2(d-c)} (\delta a - \delta c),
\label{Eq:LinEv2}\\
\delta\Phi_{tt} - e^{d-a-2c}\partial_x\left( e^{d-a+2c}\delta\Phi_x \right)
 &=& \Phi_x e^{2(d-a)} \partial_x(\delta d - \delta a + 2\delta c).
\label{Eq:LinEv3}
\end{eqnarray}

Now we are ready to obtain the master equation for $\Psi$. For
simplicity, we choose a gauge in which $\delta c = 0$ that is allowed
since we are only considering points away from the throat. In this
gauge, $\Psi = e^c\delta\Phi$ and equation~(\ref{Eq:FirstIntegral})
reduces to $2c_x\delta a = \kappa\Phi_x\delta\Phi - 2\sigma
e^{a-d-2c}$. Equation~(\ref{Eq:LinEv2}) then yields
$c_x^2\partial_x(\delta d - \delta a) = 2c_x e^{2(a-c)}\delta a =
\kappa\,e^{2(a-c)}\Phi_x\delta\Phi - 2\sigma e^{3a-d-4c}$ allowing us
to re-express the right-hand side of equation~(\ref{Eq:LinEv3}) in
terms of $\delta\Phi$. The result is the master equation
\begin{equation}
\Psi_{tt} - e^{d-a}\partial_x\left( e^{d-a}\Psi_x \right) + V(x)\Psi = Q(x),
\label{Eq:MasterPsi}
\end{equation}
with the potential
\begin{equation}
V(x) = e^{2(d-c)} - e^{2(d-a)} c_x^2 
 - \kappa e^{2(d-c)}\left( \frac{\Phi_x}{c_x} \right)^2,
\end{equation}
and the forcing term
\begin{equation}
Q(x) = -2\sigma\frac{\Phi_x}{c_x^2} e^{d+a-3c}.
\end{equation}
Since $\Psi$ has a gauge-invariant meaning, this equation holds for
arbitrary gauges, and not only gauges for which $\delta c = 0$.

For the static wormhole solution
(\ref{Eq:StaticWormHole1},\ref{Eq:StaticWormHole2}) we have
\begin{eqnarray}
e^{d-a} &=& e^{2\gamma_1\arctan(x/b) + 2\gamma_0},
\nonumber\\
V(x) &=& \frac{e^{4\gamma_1\arctan(x/b) + 4\gamma_0}}{x^2 + b^2}\left[
 1 - \frac{(x - \gamma_1 b)^2}{x^2 + b^2} 
 + \frac{2(1+\gamma_1^2)b^2}{(x - \gamma_1 b)^2} \right],
\nonumber\\
Q(x) &=& -2\sigma\Phi_1 b\,\frac{e^{3\gamma_1\arctan(x/b)-3\gamma_0}}
  {\sqrt{x^2+b^2}(x - \gamma_1 b)^2}\; .
\nonumber
\end{eqnarray}
We see explicitly that the potential $V$ diverges at the throat $x =
\gamma_1 b$. Therefore, the master equation (\ref{Eq:MasterPsi})
requires $\Psi$ to decay sufficiently fast to zero as $x\to \gamma_1
b$ in order to be meaningful. If $\delta c = 0$ everywhere this in
turn requires $\delta\Phi$ to be zero at the throat. As an example,
the potential for the zero mass wormhole simplifies to $V(x) =
(b/x)^2 (3x^2 + 2b^2)/(x^2 + b^2)^2$ and is positive away from the
throat\footnote{The positivity of $V$ led to the claim in \cite{cA02}
that there are no unstable, radial modes.}. However, the two
asymptotic regions are separated from each other by the infinite
potential wall at $x=0$ which acts like a two-sided mirror. Therefore,
it follows that with this mirror in place perturbations described by
the field $\Psi$ cannot grow in time. On the other hand, there is no
physical reason for restricting ourselves to perturbations with zero
$\Psi$ at the throat. In the next subsection we show that
equation~(\ref{Eq:MasterPsi}) can be transformed into an equation that is
everywhere regular and that does not require $\Psi$ to be zero at the
throat.

Before we proceed, we analyze the interpretation of the integration
constant $\sigma$. For this we notice that the linearization of the
static expressions (\ref{Eq:StaticWormHole1},\ref{Eq:StaticWormHole2})
with respect to the parameters $b$, $\gamma_0$ and $\gamma_1$ must
satisfy the master equation (\ref{Eq:MasterPsi}) for some value of
$\sigma$. Computing the corresponding variation, we find
\begin{equation}
\Psi = -\frac{\Phi_1 e^c}{x - \gamma_1 b}\left( e^{\gamma_0}\delta B
 - \frac{b + \gamma_1 x}{1 + \gamma_1^2}\,\arctan(x/b) 
 \delta\gamma_1 \right),
\label{Eq:StaticVariation}
\end{equation}
where $B = b e^{-\gamma_0}$. By inserting this expression into
equation~(\ref{Eq:MasterPsi}) we find that
\begin{equation}
\sigma = B\delta\gamma_1 + \gamma_1\delta B = \delta(B\gamma_1).
\end{equation}
Therefore, the integration constant $\sigma$ describes static
perturbations which change the value of the product of the asymptotic
masses, $m_\infty m_{-\infty} = -(B\gamma_1)^2$. Since any
perturbation can be written as the sum of such a static perturbation
plus a perturbation with $\delta(B\gamma_1) = 0$, we can assume that
$\sigma = 0$ for the following. The master equation
(\ref{Eq:MasterPsi}) then still admits the static solution
\begin{equation}
\Psi_0 = \frac{e^c}{x - \gamma_1 b}\left( 1 + \frac{\gamma_1}{b}
  \frac{b + \gamma_1 x}{1 + \gamma_1^2}\,\arctan(x/b) \right),
\label{Eq:PsiZero}
\end{equation}
which is obtained from (\ref{Eq:StaticVariation}) with
$\delta(\gamma_1 B) = 0$ and $\delta B = -e^{-\gamma_0}/\Phi_1$. The
existence of this solution plays an important role in the next
subsection.

\subsection{Transformation to a regular equation}
\label{Sec:LinStabMasterNew}

Here, we transform the singular master equation (\ref{Eq:MasterPsi})
into a regular one. For this, and for notational simplicity, we first
rescale the coordinates $t$ and $x$ such that $b=1$ and $\gamma_0 =
0$. Next, we define the two differential operators
\begin{displaymath}
{\cal A} := \partial - \frac{\partial\Psi_0}{\Psi_0}\; ,\qquad
{\cal A}^\dagger := -\partial - \frac{\partial\Psi_0}{\Psi_0}\; ,
\end{displaymath}
where $\partial := e^{2\gamma_1\arctan(x)}\partial_x$ and $\Psi_0$ is
the particular solution given in equation~(\ref{Eq:PsiZero}). Since ${\cal
A}^\dagger {\cal A} = -\partial^2 + (\partial^2\Psi_0)/\Psi_0 =
-\partial^2 + V\Psi_0$ we can rewrite the master equation
(\ref{Eq:MasterPsi}) as
\begin{equation}
\left( \partial_t^2 + {\cal A}^\dagger {\cal A} \right)\Psi = 0.
\end{equation}
Applying the operator ${\cal A}$ on both sides of this equation, we
find that the quantity $\chi := {\cal A}\Psi$ satisfies
\begin{displaymath}
\left( \partial_t^2 + {\cal A} {\cal A}^\dagger \right)\chi = 0,
\end{displaymath}
where
\begin{displaymath}
{\cal A} {\cal A}^\dagger 
 = -\partial^2 + \frac{\partial^2(\Psi_0^{-1})}{\Psi_0^{-1}}
 = -\partial^2 - V + 2\left( \frac{\partial\Psi_0}{\Psi_0} \right)^2.
\end{displaymath}
Therefore, $\chi$ satisfies the transformed equation
\begin{equation}
\chi_{tt} - \partial^2\chi + W(x)\chi = 0
\label{Eq:MasterChi}
\end{equation}
with the transformed potential $W = -V + 2\left(
\frac{\partial\Psi_0}{\Psi_0} \right)^2$. Explicitly, we have
\begin{displaymath}
\Psi_0(x) = \frac{\sqrt{1+x^2} e^{-\gamma_1\arctan(x)}}{x-\gamma_1}\, F(x),
\qquad F(x) := 1 + \frac{\gamma_1}{1 + \gamma_1^2}(1 + \gamma_1 x)\arctan(x).
\end{displaymath}
It is not difficult to prove that the function $F$ is strictly
positive. A short computation gives
\begin{equation}
\frac{\partial\Psi_0}{\Psi_0} = e^{2\gamma_1\arctan(x)}\left[
\frac{F_x}{F} + \frac{x-\gamma_1}{1+x^2} - \frac{1}{x-\gamma_1} \right].
\end{equation}
Using this, a lengthy calculation yields the following expression for $W$:
\begin{equation}
W(x) = e^{4\gamma_1\arctan(x)}\left[ 
 -\frac{3}{1+x^2} + 3\frac{(x-\gamma_1)^2}{(1+x^2)^2} 
 + 2\left( \frac{F_x}{F} \right)^2 - \frac{4\gamma_1}{1+x^2}\frac{F_x}{F}
 + \frac{4\gamma_1}{1+\gamma_1^2}\frac{x-\gamma_1}{F(1+x^2)^2} \right],
\end{equation}
where
\begin{displaymath}
F_x = \frac{\gamma_1}{1+\gamma_1^2}\left[ \frac{1+\gamma_1 x}{1+x^2}
 + \gamma_1\arctan(x) \right], \qquad
F = 1 + \frac{\gamma_1}{1 + \gamma_1^2}(1 + \gamma_1 x)\arctan(x).
\end{displaymath}
For $\gamma_1=0$ the new potential reduces to the simple expression
$W(x) = -3(1 + x^2)^{-2}$ which is strictly negative. For
$\gamma_1\neq 0$ we have
\begin{displaymath}
W(\gamma_1) = e^{4\gamma_1\arctan(\gamma_1)}\left[
 -\frac{3}{1+\gamma_1^2} - \frac{2\gamma_1^2}{(1+\gamma_1^2)^2} \right]
 < 0
\end{displaymath}
and $W(x) = e^{\pm 2\pi\gamma_1} \left[ 2x^{-2} + O(x^{-3}) \right]$
for $x \to \pm\infty$. Therefore, $W(x)$ is positive for large $|x|$
and negative near the throat $x = \gamma_1$. In the next subsection we
prove that the new master equation (\ref{Eq:MasterChi}) possesses a
unique unstable mode of the form
\begin{displaymath}
\chi(t,x) = e^{\beta t}\chi_0(x),
\end{displaymath}
where $\chi_0$ is the ground state of the Schr\"odinger operator $H :=
{\cal A}{\cal A}^\dagger = -\partial^2 + W$ with negative energy $E_0
= -\beta^2$. Notice that $\chi_0(x)$ decays exponentially to zero as
$|x|\to\infty$.

\subsection{Existence of a bound state with negative energy}
\label{Sec:LinStabBoundState}

Here, we establish the existence of a bound state with negative energy
of the Schr\"odinger operator $H := {\cal A}{\cal A}^\dagger$. For
this, we first introduce the new coordinate
\begin{displaymath}
\rho(x) := \int\limits_0^x e^{-2\gamma_1\arctan(y)} dy
\end{displaymath}
which monotonically increases from $-\infty$ to $+\infty$ and
satisfies $\partial = e^{2\gamma_1\arctan(x)}\partial_x =
\partial_\rho$ and $\lim\limits_{x\to\pm\infty} \rho/x =
\exp(\mp\gamma_1\pi)$. Since $W$ is real and bounded, $H =
-\partial_\rho^2 + W(\rho)$ defines a self-adjoint operator on the
Hilbert space $L^2(\Real,d\rho)$ with domain consisting of the dense
subspace $D(H) := H^2(\Real,d\rho)$ of functions whose zeroth, first
and second order derivatives are square-integrable on $\Real$. Let us
assume first that $\gamma_1 \neq 0$. In this case the function
\begin{equation}
\frac{1}{\Psi_0} 
 = \frac{x-\gamma_1}{\sqrt{1+x^2} F(x)}\, e^{\gamma_1\arctan(x)}
\label{Eq:ZeroMode}
\end{equation}
is an eigenfunction of $H$ with zero eigenvalue since $W =
\Psi_0\partial_\rho^2(\Psi_0^{-1})$ and since $F$ grows linearly in
$|x|$ for large $|x|$. Additionally, this zero mode has precisely one
zero, namely at the throat $x = \gamma_1$. It follows from the nodal
theorem\footnote{See, for instance Ref. \cite{CourantHilbert-Book}.}
that $H$ possesses exactly one bound state $\chi_0$ with
negative energy $E_0 = -\beta^2 < 0$. Its value will be computed in
the next section via a numerical shooting method.

For $\gamma_1 = 0$ the function $1/\Psi_0 = x/\sqrt{1+x^2}$ is not an
eigenfunction of $H$ anymore because it does not decay to zero as
$|x|\to\infty$. Nevertheless, we may still expect the existence of a
unique bound state with negative energy by continuity. In order to
obtain an upper bound for the ground state energy $E_0$ for the case
$\gamma_1=0$ where $W = -3/(1+x^2)^2$, we use the
Rayleigh-Ritz\footnote{See, for instance
  Ref. \cite{CourantHilbert-Book}.}  variational principle,
\begin{equation}
E_0 = \inf\limits_{\Psi\in D(H)\setminus \{ 0 \}} 
\frac{ (\Psi,H\Psi) }{(\Psi,\Psi)} \; ,
\end{equation}
where $(\cdot\, ,\cdot)$ denotes the scalar product of
$L^2(\Real,d\rho)$. Testing with the particular functions $\Psi_K(x) =
(1 + x^2)^{-K/2}$, $K > 1/2$, gives
\begin{displaymath}
(\Psi_K,\Psi_K) = 2I_K, \qquad
(\Psi_K, H\Psi_K) = 2\left[ K^2 I_{K+1} - (K^2+3) I_{K+2} \right],
\end{displaymath}
where the integral
\begin{displaymath}
I_K := \int\limits_0^\infty \frac{dx}{(1 + x^2)^K}\; ,\qquad
K > \frac{1}{2}\; ,
\end{displaymath}
satisfies the recursion relation
\begin{displaymath}
I_{K+1} = \frac{2K-1}{2K} I_K\; ,\qquad K > \frac{1}{2}\; .
\end{displaymath}
Therefore, we obtain the bound
\begin{equation}
E_0 \leq \frac{1}{4} f(K), \qquad
f(K) := \frac{(2K-1)(K^2 - 6K - 3)}{K(K+1)}\; ,\qquad K > \frac{1}{2}\; .
\end{equation}
The function $f$ is negative for $1/2 < K < 3 + 2\sqrt{3}$ and
positive for $K > 3 + 2\sqrt{3}$. Its minimum lies near $K=2$ for
which $f(K) = -11/2$. Since the minimum of $W$ is $-3$ we obtain the
estimate $-3 \leq E_0 \leq -11/8 = -1.375$ for the ground state
energy. This estimate is used as a starting point for a numerical
shooting method described in Sec. \ref{Sec:TimeScale}.

\subsection{Existence of an exponentially growing mode}
\label{Sec:LinStabUnstable}

To prove that the unstable mode found in the previous subsection gives
rise to an exponentially growing regular solution of the linearized
field equations we still need to show that there exist perturbation
amplitudes $\delta d$, $\delta a$, $\delta c$ and $\delta\Phi$ which
satisfy the linearized field equations, are everywhere regular and
grow exponentially in time. Here, we show that such a solution exists
indeed.

In order to do so we first notice that $\Psi = {\cal A}^\dagger\chi$
satisfies the master equation (\ref{Eq:MasterPsi}). Next, we choose the
infinitesimal coordinates such that $\delta\Phi = 0$. This is possible
because $\Phi_x = \Phi_1/(1+x^2)\neq 0$ for all $x\in\Real$. In this
gauge, $\Psi = -\Phi_1 e^c\delta c/(x-\gamma_1)$, and so we obtain
\begin{equation}
e^c\delta c = -\frac{x-\gamma_1}{\Phi_1}\Psi 
 = \frac{x-\gamma_1}{\Phi_1}
   \left( \partial + \frac{\partial\Psi_0}{\Psi_0} \right)\chi
 = \frac{e^{2\gamma_1\arctan(x)}}{\Phi_1}\left[
 (x-\gamma_1)\partial_x + (x-\gamma_1)\frac{F_x}{F} 
 + \frac{(x-\gamma_1)^2}{1+x^2} - 1 \right]\chi,
\label{Eq:deltac}
\end{equation}
which is regular everywhere and decays exponentially fast to zero as
$|x|\to\infty$. We may obtain $\delta d$ and $\delta a$ from this
using the perturbation equations (\ref{Eq:FirstIntegral}) and
(\ref{Eq:LinEv3}) which, in the gauge $\delta\Phi=0$ yield
\begin{eqnarray}
c_x\delta a &=& \delta c_x + (c_x - d_x)\delta c,
\nonumber\\
\delta d_x &=& \delta a_x - 2\delta c_x\; .
\nonumber
\end{eqnarray}
A careful calculation using the fact that $\chi$ satisfies
$(\partial^2 - W)\chi = \beta^2\chi$ and the identity
\begin{equation}
\frac{F_x}{F} - \frac{\gamma_1}{1+x^2}
 = \frac{1}{F}\frac{\gamma_1^2(x-\gamma_1)}{(1+\gamma_1^2)(1+x^2)}
\left[ 1 + x\arctan(x) \right]
\label{Eq:Identity}
\end{equation}
yields
\begin{eqnarray}
\delta a &=& \left[
 1 + \frac{\gamma_1^2}{1+\gamma_1^2}\frac{1+x\arctan(x)}{F} \right]\delta c
 + \frac{\sqrt{1+x^2}}{\Phi_1} e^{-\gamma_1\arctan(x)}\beta^2\chi,
\label{Eq:deltaa}\\
\delta d &=& \left[
 -1 + \frac{\gamma_1^2}{1+\gamma_1^2}\frac{1+x\arctan(x)}{F} \right]\delta c
 + \frac{\sqrt{1+x^2}}{\Phi_1} e^{-\gamma_1\arctan(x)}\beta^2\chi
 + h(t),
\label{Eq:deltad}
\end{eqnarray}
where $h(t)$ is an arbitrary constant which depends on $t$ only and
could be eliminated by an infinitesimal coordinate transformation. It
can be checked that the expressions
(\ref{Eq:deltac},\ref{Eq:deltaa},\ref{Eq:deltad}) also satisfy the
evolution equations (\ref{Eq:LinEv2}) and the linearization of
equation~(\ref{Eq:Ev1}) with $\delta\Phi=0$.

Therefore, we obtain an exponentially growing solution of the
linearized field equations which is everywhere regular and decays
exponentially to zero for $|x| \to \infty$. This proves the linear
instability of static, spherically symmetric wormholes supported by a
massless ghost scalar field. Their nonlinear evolution are studied in
a subsequent paper.

%%%%%%%%%%%%%%%%%%%%%%%%%%%%%%%%%%%%%%%%%%%%%%%%%%%%%%%%%%%%%%%%%%%%%%%
\section{Time scale of the instability}
\label{Sec:TimeScale}
%%%%%%%%%%%%%%%%%%%%%%%%%%%%%%%%%%%%%%%%%%%%%%%%%%%%%%%%%%%%%%%%%%%%%%%

In this section we compute the ground state energy $E_0 = -\beta^2$ of
the Schr\"odinger operator $H$ via a numerical shooting method. For
this, we first multiply the differential equation $H\chi =
-\beta^2\chi$ on both sides by the factor
$e^{-4\gamma_1\arctan(\gamma_1)}$ and obtain
\begin{equation}
-\left[ a(x)\partial_x \right]^2\chi 
 + \left[ \bar{\beta}^2 + \bar{W}(x) \right]\chi = 0,
\label{Eq:DiffEq}
\end{equation}
where $a(x) = e^{2\gamma_1[\arctan(x) - \arctan(\gamma_1)]}$,
$\bar{\beta} = e^{-2\gamma_1\arctan(\gamma_1)}\beta$ and
\begin{displaymath}
\bar{W}(x) = a(x)^2\left[ -\frac{3}{1+x^2} + 3\frac{(x-\gamma_1)^2}{(1+x^2)^2} 
 + 2\left( \frac{F_x}{F} \right)^2 - \frac{4\gamma_1}{1+x^2}\frac{F_x}{F}
 + \frac{4\gamma_1}{1+\gamma_1^2}\frac{x-\gamma_1}{F(1+x^2)^2} \right].
\end{displaymath}
For $\gamma_1=0$ the potential is strictly negative. For $\gamma_1 >
0$ it is convenient to rewrite the potential as
\begin{displaymath}
\bar{W}(x) = a(x)^2\left\{  
\frac{(\gamma_1^2 - 3 - 6\gamma_1 x) l(x)^2 - 2\gamma_1^{-2}}
     {(1+x^2)^2 l(x)^2}
 + 2\left[ \frac{1}{F}\frac{\gamma_1^2(x-\gamma_1)}{(1+\gamma_1^2)(1+x^2)}
 l(x) + \frac{\gamma_1^{-1}}{(1+x^2) l(x)} \right]^2
 \right\},
\end{displaymath}
where $l(x) := 1 + x\arctan(x) \geq 1$ and where we have used the
identity (\ref{Eq:Identity}). Since the second term inside the curly
brackets is a perfect square we see from this representation that
$\bar{W}(x)$ is positive for all values of $x$ small enough such that
\begin{equation}
(6\gamma_1 x + 3 - \gamma_1^2)l(x)^2 < -2\gamma_1^{-2}.
\label{Eq:Inequality}
\end{equation}
Therefore, for small $x$, $\bar{\beta}^2 + \bar{W}(x) > 0$ which means
that $\chi$ cannot have a local maximum (minimum) if $\chi > 0$ ($\chi
< 0$). For large $x$, on the other hand, $\bar{W}(x) =
e^{2\gamma_1(\pi - 2\arctan(\gamma_1))}\left[ 2x^{-2} + O(x^{-3})
\right]$ decays like $1/x^2$ and so the normalizable solutions behave
like
\begin{equation}
\chi(x) \approx e^{-\bar{\beta}\bar{\rho}(x)},
\label{Eq:AsymSol}
\end{equation}
for large $x$, where $\bar{\rho}(x) = \int_0^x\frac{dy}{a(y)}$.

Our shooting procedure consists in integrating the differential
equation (\ref{Eq:DiffEq}) starting with the asymptotic solution
(\ref{Eq:AsymSol}) at some large value of $x$ and integrating
numerically towards small values of $x$. The integration is stopped as
soon as the inequality (\ref{Eq:Inequality}) is satisfied and
$\chi\cdot\chi_x < 0$. If this is the case, the solution must diverge
as $x\to -\infty$ since $|\chi|$ cannot have a local maxima in the
region where the inequality (\ref{Eq:Inequality}) is
satisfied. Therefore, we look for the value of $\bar{\beta}$ for which
$\chi\to 0$ as $x\to -\infty$.

\begin{figure}[ht]
\centerline{\resizebox{9cm}{!}{\includegraphics{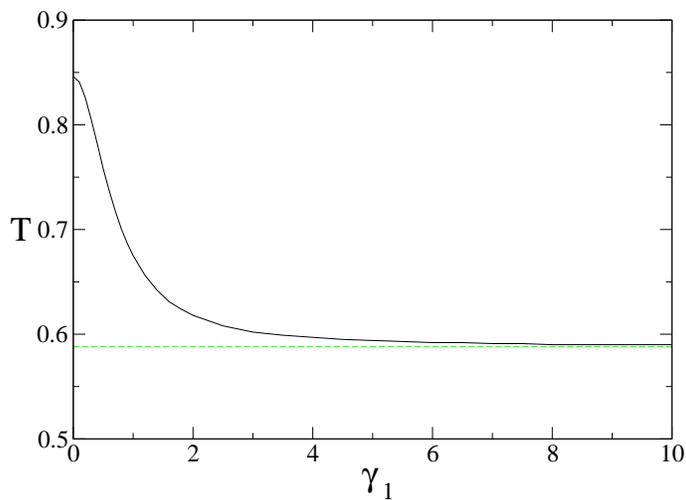}}}
\vspace{0.3cm}
\caption{The dimensionless time scale $T:=\tau_{unstable}/r_{throat}$
vs. $\gamma_1$. This plot and the values in the previous table
strongly suggest that $T$ converges to the value of $0.590$ as
$\gamma_1\to\infty$. The dashed line represents the value for $T$ at
$\gamma_1\to\infty$ which is computed in the next subsection.}
\label{Fig:TimeScale}
\end{figure}

For the numerical integration we compactify the domain by means of the
transformation $x = \tan(\pi z/2)$ which maps $z\in (-1,1)$ onto
$x\in\Real$ and use a fourth order Runge-Kutta integrator with fixed
step size. We start with the massless case $\gamma_1 = 0$ where the
estimates in the previous section provide the bound $\sqrt{11/8} \leq
\bar{\beta} \leq \sqrt{3}$ for the parameter $\bar{\beta}$. For
$\bar{\beta} = \sqrt{11/8}$ the final value for $\chi$ at small $x$ is
negative, for $\bar{\beta} = \sqrt{3}$ it is positive. The optimal
value for $\bar{\beta}$ is obtained by bisecting the interval
$[\sqrt{11/8},\sqrt{3}]$. Then, we increment the value of $\gamma_1$
by a small amount and repeat the above shooting method, finding the
optimal value for $\bar{\beta}$ for the new value of
$\gamma_1$. Continuing this way we obtain the results summarized in
table \ref{Tab:TimeScale} and figure \ref{Fig:TimeScale}, where we
show the associated time scale in terms of proper time at the throat
of the background solution, $\tau = e^{\gamma_1\arctan(\gamma_1)} t$,
and units of the areal radius of the throat, $r_{throat} =
b\sqrt{1+\gamma_1^2} e^{-\gamma_1\arctan(\gamma_1)}$. This time scale
is given by
\begin{displaymath}
\tau_{unstable} = \frac{e^{2\gamma_1\arctan(\gamma_1)}}{\sqrt{1+\gamma_1^2}}
\frac{r_{throat}}{\beta} 
 = \frac{r_{throat}}{\sqrt{1+\gamma_1^2}\,\bar{\beta}}\; .
\end{displaymath}
As is apparent from the results in table \ref{Tab:TimeScale} and
figure \ref{Fig:TimeScale}, the time scale converges to a value of
$0.590 r_{throat}$ for large values of $\gamma_1$. The limit
$\gamma_1\to\infty$ is analyzed next.

\begin{table}[h]
\center
\begin{tabular}{|l||c|c|c|c|c|c|c|c|c|c|c|c|c|c|c|c|}\hline
 $\gamma_1$ & 
$0$     & $0.1$   & $0.2$   & $0.3$   & $0.4$   & $0.5$   & 
$0.6$   & $0.7$   & $0.8$   & $0.9$   & $1.0$   & $1.2$   &
$1.4$   & $1.6$   & $1.8$   & $2.0$ \\ 
\hline
$T$ &
$0.846$ & $0.841$ & $0.826$ & $0.805$ & $0.782$ & $0.758$ &  
$0.737$ & $0.718$ & $0.701$ & $0.687$ & $0.675$ & $0.656$ &
$0.642$ & $0.631$ & $0.624$ & $0.618$ \\
\hline
\end{tabular}

\begin{tabular}{|l||c|c|c|c|c|c|c|c|c|c|c|c|c|c|c|c|}\hline
 $\gamma_1$ & 
$2.5$   & $3.0$   & $3.5$   & $4.0$   & $4.5$   & $5.0$   & 
$5.5$   & $6.0$   & $6.5$   & $7.0$   & $7.5$   & $8.0$   &
$8.5$   & $9.0$   & $9.5$   & $10.0$ \\ 
\hline
$T$ &
$0.608$ & $0.602$ & $0.599$ & $0.597$ & $0.595$ & $0.594$ &  
$0.593$ & $0.592$ & $0.592$ & $0.591$ & $0.591$ & $0.590$ &
$0.590$ & $0.590$ & $0.590$ & $0.590$ \\
\hline
\end{tabular}
\caption{Numerical values for the dimensionless time scale
$T:=\tau_{unstable}/r_{throat}$ for different values of $\gamma_1$.}
\label{Tab:TimeScale}
\end{table}

\vspace{0.4cm}

\subsection{The large $\gamma_1$ limit}

In order to understand the behavior of the time scale for large
$\gamma_1$ we first notice that the inequality (\ref{Eq:Inequality})
is satisfied for $x < \gamma_1/6 - 1/(2\gamma_1) - 1/(3\gamma_1^3) =:
R(\gamma_1)$. Therefore, the potential $\bar{W}(x)$ is strictly
positive for all $x\in (-\infty,R(\gamma_1))$ and the solution grows
at least exponentially fast with increasing $x$ on this interval.
Since $R(\gamma_1) \to \infty$ as $\gamma_1\to\infty$, the solution
must vanish for each fixed $x$ as $\gamma_1\to\infty$ if it is
required to have a fixed, finite value at $x = R(\gamma_1)$. For this
reason, we first replace the coordinate $x$ by $z := x/\gamma_1$ in
the following, and then take the limit $\gamma_1\to\infty$.

In terms of the new coordinate $z$ the differential equation
(\ref{Eq:DiffEq}) is transformed into $-[\hat{a}(z)\partial_z]^2\chi +
[\hat{\beta}^2 + \hat{W}]\chi = 0$, where $\hat{a}(z) =
e^{2\gamma_1[\arctan(\gamma_1 z) - \arctan(\gamma_1)]}$, $\hat{\beta} =
\gamma_1\bar{\beta}$ and
\begin{displaymath}
\hat{W}(z) = \hat{a}(z)^2\left[  
 -\frac{3\gamma_1^2}{1+\gamma_1^2 z^2} 
 + 3\frac{\gamma_1^4(z-1)^2}{(1+\gamma_1^2 z^2)^2} 
 + 2\left( \frac{F_z}{F} \right)^2 
 - \frac{4\gamma_1^2}{1+\gamma_1^2 z^2}\frac{F_z}{F}
 + \frac{4\gamma_1^2}{1+\gamma_1^2}
   \frac{\gamma_1^2(z-1)}{F(1+\gamma_1^2 z^2)^2} \right],
\end{displaymath}
where $F_z = \gamma_1^2(1+\gamma_1^2)^{-1} [ (1+\gamma_1^2
 z)/(1+\gamma_1^2 z^2) + \gamma_1\arctan(\gamma_1 z)]$ and $F = 1 +
 \gamma_1(1+\gamma_1^2)^{-1}(1 + \gamma_1^2 z)\arctan(\gamma_1
 z)$. From the above, $\hat{W}(z) > 0$ if $z < 1/6 - 1/(2\gamma_1^2) -
 1/(3\gamma_1^4)$. Now let $z > 0$ be fixed. In the limit
 $\gamma_1\to\infty$ we have
\begin{displaymath}
\frac{F}{\gamma_1} \to \frac{\pi}{2}\; z, \qquad
\frac{F_z}{\gamma_1} \to \frac{\pi}{2}\; ,
\end{displaymath}
and
\begin{displaymath}
\gamma_1[\arctan(\gamma_1 z) - \arctan(\gamma_1)] 
 = \gamma_1\arctan\left( \frac{\gamma_1(z-1)}{1 + \gamma_1^2 z} \right)
 \to \frac{z-1}{z}\; .
\end{displaymath}
Therefore, the differential equation (\ref{Eq:DiffEq}) converges
pointwise to
\begin{equation}
-\left( e^{2(1-z^{-1})}\partial_z \right)^2 \chi
 +  e^{4(1-z^{-1})}\left[ 2 z^{-2} - 10 z^{-3} + 3 z^{-4} \right]\chi
 = -\hat{\beta}^2\chi, \qquad z > 0.
\label{Eq:DiffEqLargeGamma}
\end{equation}
The operator on the left-hand side is a formally self-adjoint operator
on the Hilbert space $L^2( (0,\infty), e^{-2(1-z^{-1})}dz)$ which is
singular at $z=0$. We obtain a zero mode of this operator by
multiplying equation~(\ref{Eq:ZeroMode}) by $\gamma_1
e^{-\gamma_1\arctan(\gamma_1)}$ and taking the limit $\gamma_1\to
\infty$. This yields
\begin{equation}
\frac{\gamma_1}{\Psi_0} e^{-\gamma_1\arctan(\gamma_1)} 
 = \frac{\gamma_1(z-1)}{\sqrt{1 + \gamma_1^2 z^2}}\frac{\gamma_1}{F}
   e^{\gamma_1[\arctan(\gamma_1 z) - \arctan(\gamma_1)]}
 \to \frac{2}{\pi}\frac{z-1}{z^2}\, e^{1 - z^{-1}}.
\label{Eq:ZeroModeLargeGamma}
\end{equation}
Indeed, it can be checked that (\ref{Eq:ZeroModeLargeGamma}) solves
the differential equation (\ref{Eq:DiffEqLargeGamma}) with
$\hat{\beta} = 0$. Furthermore, this solution decays like $1/z$ for
large $z$ and as $z \to 0$ it decays rapidly to zero. Since it has
exactly one zero, it follows again from the nodal theorem the
existence of a unique bound state with negative energy $\hat{E}_0 =
-\hat{\beta}^2$. A numerical shooting procedure similar to the one
described above which starts at some large value of $z$, where
$\chi(z) \approx e^{-\hat{\beta}\hat{\rho}(z)}$, $\hat{\rho}(z) =
\int_1^z e^{-2(1-\zeta^{-1})} d\zeta$, yields $1/\hat{\beta} = 0.588$.
Since on the other hand $\tau_{unstable}/r_{throat} =
1/(\sqrt{1+\gamma_1^2}\bar{\beta}) = 1/\hat{\beta}$ in the limit
$\gamma_1\to\infty$ this matches well the asymptotic value obtained in
table~\ref{Tab:TimeScale}.

%%%%%%%%%%%%%%%%%%%%%%%%%%%%%%%%%%%%%%%%%%%%%%%%%%%%%%%%%%%%%%%%%%%%%%%
\section{Conclusions}
\label{Sec:Conclusions}
%%%%%%%%%%%%%%%%%%%%%%%%%%%%%%%%%%%%%%%%%%%%%%%%%%%%%%%%%%%%%%%%%%%%%%%

In this article we analyzed the stability of static, spherically
symmetric general relativistic wormhole solutions sourced by a
massless ghost scalar field. We found that all these solutions are
unstable with respect to linear fluctuations of the metric and the
scalar field. Furthermore we showed that the time scale associated to
this instability is of the order of the areal radius of the throat
divided by the speed of light, which is of the order of a few
microseconds for a throat of radius of the order of kilometers.
Therefore, the instability we found is likely to introduce a rapid
growth or collapse of the throat which eventually may destroy the
wormhole. In a subsequent paper \cite{jGfGoS-inprep2} we follow the
nonlinear evolution of the unstable mode by numerical means and show
that the wormhole either expands rapidly or collapses to a
Schwarzschild black hole.

Our conclusion about the instability of the zero mass wormhole is
different from the results presented in Ref. \cite{cA02}. As explained
in Sec. \ref{Sec:LinStabMasterPhi} this is due to the fact that the
results in \cite{cA02} only apply to a restricted class of
perturbations which are required to vanish at the throat. As shown in
this article, both the massless and massive wormholes are linearly
unstable for the more general class of spherically symmetric
perturbations which do not necessarily vanish at the throat.

The question arises whether or not there exist wormhole solutions
different from the ones considered in this article which are linearly
stable. Among the potential options are: i) wormholes with a single
asymptotic end obtained by removing two holes from $\Real^3$ and
connecting them to each other by a handle, ii) rotating wormholes,
iii) wormholes which are supported by a different kind of matter
field, iv) wormholes in modified or higher-dimensional theories of
gravity. Regarding option i), it is conceivable that a change in
topology affects our instability result. Indeed, since the linear
stability of a given solution is directly related to the spectral
properties of the linear operator appearing in the perturbation
equation it depends on the global structure of the wormhole, and not
only on local properties of the throat. Concerning option ii), the
possibility that the rotation could stabilize a wormhole has been
explored in \cite{tMdN06} where an exact solution describing the inner
region of a rotating wormhole is constructed. However, based on our
instability results, our expectation is that the centrifugal force due
to rotation only shifts the equilibrium between the attractive
gravitational force and the repulsive force of the ghost scalar field,
but does not change its stability. At least we expected that by
continuity, slowly rotating wormholes are unstable. Regarding options
iii) and iv) there is already a vast amount of results in the
literature. To mention only a few examples, it has been shown that
static spherically symmetric wormholes in scalar-tensor theories of
gravity are linearly unstable \cite{kBsG04,kBaS07}. On the other hand,
a class of wormhole solutions which are supported by phantom energy
have recently been constructed in \cite{sS05,fL05} and shown to be
stable {\em with respect to perturbations inside this class}
\cite{fL05b}.

%%%%%%%%%%%%%%%%%%%%%%%%%%%
%%%   ACKNOWLEDGMENTS   %%%
%%%%%%%%%%%%%%%%%%%%%%%%%%%

\acknowledgments

It is a pleasure to thank Ulises Nucamendi and Thomas Zannias for many
stimulating discussions. This work was supported in part by grants CIC
4.9, 4.19 and 4.23 to Universidad Michoacana, PROMEP UMICH-PTC-121,
UMICH-PTC-195, UMICH-PTC-210 and UMICH-CA-22 from SEP Mexico and
CONACyT grant numbers 61173, 79601 and 79995.

%%%%%%%%%%%%%%%%%%%%%%%%%%%%%%%%%%%%%%%%%%%%%%%%%%%%%%%%%%%%%%
% Create the reference section using BibTeX:
\bibliographystyle{unsrt}
\bibliography{refs}

\begin{thebibliography}{10}

\bibitem{mMkT88}
M.S. Morris and K.S. Thorne.
\newblock Wormholes in spacetime and their use for interstellar travel: {A}
  tool for teaching general relativity.
\newblock {\em Am. J. Phys.}, 56:395--412, 1988.

\bibitem{mMkTyU88}
M.S. Morris, K.S. Thorne, and U.~Yurtsever.
\newblock Wormholes, time machines, and the weak energy condition.
\newblock {\em Phys. Rev. Lett.}, 61:1446--1449, 1988.

\bibitem{vFiN90}
V.P. Frolov and I.D. Novikov.
\newblock Physical effects in wormholes and time machines.
\newblock {\em Phys. Rev. D}, 42:1057--1065, 1990.

\bibitem{jFkSdW93}
J.L. Friedman, K.~Schleich, and D.M. Witt.
\newblock Topological censorship.
\newblock {\em Phys. Rev. Lett.}, 71:1486--1489, 1993.

\bibitem{mVsKnD03}
M.~Visser, S.~Kar, and N.~Dadhich.
\newblock Traversable wormholes with arbitrarily small energy condition
  violations.
\newblock {\em Phys. Rev. Lett.}, 90:201102, 2003.

\bibitem{kBsG02}
K.A. Bronnikov and S.~Grinyok.
\newblock Instability of wormholes with a nonminimally coupled scalar field.
\newblock {\em Grav. Cosmol.}, 7:297--300, 2001.

\bibitem{kBsG04}
K.A. Bronnikov and S.V. Grinyok.
\newblock Conformal continuations and wormhole instability in scalar-tensor
  gravity.
\newblock {\em Grav. Cosmol.}, 10:237--, 2004.

\bibitem{kBaS07}
K.A. Bronnikov and A.A. Starobinsky.
\newblock No realistic wormholes from ghost-free scalar-tensor phantom dark
  energy.
\newblock {\em JETP Lett.}, 85:1--5, 2007.

\bibitem{cA02}
C.~Armend\'ariz-Pic\'on.
\newblock On a class of stable, traversable {L}orentzian wormholes in classical
  general relativity.
\newblock {\em Phys. Rev. D}, 65:104010, 2002.

\bibitem{hSsH02}
Hisa aki Shinkai and Sean~A. Hayward.
\newblock Fate of the first traversible wormhole: {B}lack hole collapse or
  inflationary expansion.
\newblock {\em Phys. Rev. D}, 66:044005, 2002.

\bibitem{sSsK04}
S.V. Sushkov and S-W. Kim.
\newblock Cosmological evolution of a ghost scalar field.
\newblock {\em Gen. Rel. Grav.}, 36:1671--1678, 2004.

\bibitem{tMdN06}
T.~Matos and D.~Nu{\~n}ez.
\newblock Rotating scalar field wormhole.
\newblock {\em Class. Quant. Grav.}, 23:4485--4496, 2006.

\bibitem{eDtZ07}
J.~Estevez Delgado and T.~Zannias.
\newblock On wormholes and black holes solutions of {E}instein gravity coupled
  to a {K}-massless scalar field.
\newblock {\em J. Phys. Conf. Ser.}, 66:012029, 2007.

\bibitem{hE73}
H.G. Ellis.
\newblock Ether flow through a drainhole: {A} particle model in general
  relativity.
\newblock {\em J. Math. Phys.}, 14:104--118, 1973.

\bibitem{kB73}
K.A. Bronnikov.
\newblock Scalar-tensor theory and scalar charge.
\newblock {\em Acta Phys. Polonica B}, 4:251--266, 1973.

\bibitem{cMdS64}
C.W. Misner and D.H. Sharp.
\newblock Relativistic equations for adiabatic, spherically symmetric
  gravitational collapse.
\newblock {\em Phys. Rev.}, 136:B571--B576, 1964.

\bibitem{oBmHnS96}
O.~Brodbeck, M.~Heusler, and N.~Straumann.
\newblock Pulsations of spherically symmetric systems in general relativity.
\newblock {\em Phys. Rev. D}, 53:765--761, 1996.

\bibitem{CourantHilbert-Book}
R.~Courant and D.~Hilbert.
\newblock {\em Methods of Mathematical Physics, Volume I}.
\newblock Wiley Classics Edition, 1989.

\bibitem{jGfGoS-inprep2}
J.~A. Gonz\'alez, F.~S. Guzm\'an, and O.~Sarbach.
\newblock Instability of wormholes supported by a ghost scalar field {II}:
  {N}onlinear evolution.
\newblock http://arxiv.org/abs/0806.13708, to appear in Class. Quantum Grav.

\bibitem{sS05}
S.V. Sushkov.
\newblock Wormholes supported by a phantom energy.
\newblock {\em Phys. Rev. D}, 71:043520, 2005.

\bibitem{fL05}
F.S.N. Lobo.
\newblock Phantom energy traversable wormholes.
\newblock {\em Phys. Rev. D}, 71:084011, 2005.

\bibitem{fL05b}
F.S.N. Lobo.
\newblock Stability of phantom wormholes.
\newblock {\em Phys. Rev. D}, 71:124022, 2005.

\end{thebibliography}
%%%%%%%%%%%%%%%%%%%%%%%%%%%%%%%%%%%%%%%%%%%%%%%%%%%%%%%%%%%%%%
\end{document}